 \documentstyle[epsf,twocolumn,prd,aps]{revtex}
\begin{document}

\draft
%
%
\newcommand{\nc}{\newcommand}
\nc{\bea}{\begin{eqnarray}}
\nc{\eea}{\end{eqnarray}}
\nc{\beq}{\begin{equation}}
\nc{\eeq}{\end{equation}}
\nc{\bi}{\begin{itemize}}
\nc{\ei}{\end{itemize}}
\nc{\la}[1]{\label{#1}}
\nc{\half}{\frac{1}{2}}
\nc{\fsky}{f_{\rm sky}}
\nc{\fwhm}{\theta_{\rm fwhm}}
\nc{\fwhmc}{\theta_{{\rm fwhm},c}}
\nc{\GeV}{\mbox{ GeV}}
\nc{\MeV}{\mbox{ MeV}}
\nc{\keV}{\mbox{ keV}}
\nc{\etal}{{\it et al.}}
\nc{\x}[1]{}
%
%

\nc{\AJ}[3]{{Astron.~J.\ }{{\bf #1}{, #2}{ (#3)}}}
\nc{\anap}[3]{{Astron.\ Astrophys.\ }{{\bf #1}{, #2}{ (#3)}}}
\nc{\ApJ}[3]{{Astrophys.~J.\ }{{\bf #1}{, #2}{ (#3)}}}
\nc{\apjl}[3]{{Astrophys.~J.\ Lett.\ }{{\bf #1}{, #2}{ (#3)}}}
\nc{\app}[3]{{Astropart.\ Phys.\ }{{\bf #1}{, #2}{ (#3)}}}
\nc{\araa}[3]{{Ann.\ Rev.\ Astron.\ Astrophys.\ }{{\bf #1}{, #2}{ (#3)}}}
\nc{\arns}[3]{{Ann.\ Rev.\ Nucl.\ Sci.\ }{{\bf #1}{, #2}{ (#3)}}}
\nc{\arnps}[3]{{Ann.\ Rev.\ Nucl.\ and Part.\ Sci.\ }{{\bf #1}{, #2}{ (#3)}}}
\nc{\MNRAS}[3]{{Mon.\ Not.\ R.\ Astron.\ Soc.\ }{{\bf #1}{, #2}{ (#3)}}}
\nc{\mpl}[3]{{Mod.\ Phys.\ Lett.\ }{{\bf #1}{, #2}{ (#3)}}}
\nc{\Nat}[3]{{Nature }{{\bf #1}{, #2}{ (#3)}}}
\nc{\ncim}[3]{{Nuov.\ Cim.\ }{{\bf #1}{, #2}{ (#3)}}}
\nc{\nast}[3]{{New Astronomy }{{\bf #1}{, #2}{ (#3)}}}
\nc{\np}[3]{{Nucl.\ Phys.\ }{{\bf #1}{, #2}{ (#3)}}}
\nc{\pr}[3]{{Phys.\ Rev.\ }{{\bf #1}{, #2}{ (#3)}}}
\nc{\PRD}[3]{{Phys.\ Rev.\ D\ }{{\bf #1}{, #2}{ (#3)}}}
\nc{\PRL}[3]{{Phys.\ Rev.\ Lett.\ }{{\bf #1}{, #2}{ (#3)}}}
\nc{\PL}[3]{{Phys.\ Lett.\ }{{\bf #1}{, #2}{ (#3)}}}
\nc{\prep}[3]{{Phys.\ Rep.\ }{{\bf #1}{, #2}{ (#3)}}}
\nc{\RMP}[3]{{Rev.\ Mod.\ Phys.\ }{{\bf #1}{, #2}{ (#3)}}}
\nc{\rpp}[3]{{Rep.\ Prog.\ Phys.\ }{{\bf #1}{, #2}{ (#3)}}}
\nc{\ibid}[3]{{\it ibid.\ }{{\bf #1}{, #2}{ (#3)}}}

\wideabs{
\title{Constraining Isocurvature Fluctuations with the 
       Planck Surveyor}

\author{Kari Enqvist\cite{mailk}}
\address{Department of Physics, University of Helsinki, and Helsinki Institute of Physics,\\
         P.O.Box 9, FIN-00014 University of Helsinki, Finland}

\author{Hannu Kurki-Suonio\cite{mailh}}
\address{Helsinki Institute of Physics,
         P.O.Box 9, FIN-00014 University of Helsinki, Finland}

\maketitle

\begin{abstract}
We consider the detection possibilities of isocurvature fluctuations
in the future CMB satellite experiments MAP and Planck
for different cosmological
reference models. We present a simultaneous 10 parameter fit
(8 for the case of open model) to determine the correlations
between the cosmological parameters, including isocurvature cold dark matter
contribution to the anisotropy.
Assuming that polarization information can be fully exploited,
we find that an isocurvature perturbation can be detected by the
Planck Surveyor if the ratio of the initial isocurvature
and adiabatic perturbation amplitude is larger than 0.07. In
the absence of polarization data, the signal from isocurvature
perturbations can be confused with tensor perturbations or early
reionization effects, and the limit is larger by almost
an order of magnitude.
\end{abstract}

}

%
%
\section{Introduction}

A second generation of  satellite experiments, MAP\cite{MAP} and 
Planck\cite{Planck}, will soon provide 
highly detailed temperature maps of the cosmic microwave background 
(CMB). From the measured spectrum of temperature fluctuations one may infer
the properties of the cosmic fluid, such as its density or the rate by
which it comoves with the expansion of the Universe, and thus determine most
cosmological parameters to an accuracy of a few
percent\cite{Jungmanetal,Bondetal,ZSS97}. In addition, the
spectrum of temperature fluctuations allows us to study and constrain physics
at very small distance scales, corresponding to energies much higher than
can be achieved in any particle accelerator in the foreseeable future. 
This is so because the features on the temperature map
depend on the nature of the
primordial density fluctuation. 

Cosmic inflation is currently the most popular 
explanation for the origin of the primordial perturbations,
and it typically predicts a
near scale invariant spectrum with Gaussian fluctuations. The perturbations are
adiabatic with the number density proportional to entropy density 
so that $\delta (n/s) = 0$.  This is so because the 
quantum fluctuations of the field responsible
for inflation, called the inflaton, give directly rise to perturbations
in the energy density of the inflaton field.

However, the inflaton may not be the only field which is subject to quantum
fluctuations during inflation. In fact, any effectively massless scalar field 
will fluctuate by virtue of the nearly constant energy density of 
the inflation era with a root mean square amplitude $H/(2\pi)$, where
$H$ is the Hubble parameter during inflation. Such fluctuations will
contribute to the perturbations in the microwave background. They may
be adiabatic, in which case it is difficult to entangle their effect from
the inflaton fluctuations. The other possibility is isocurvature fluctuations,
which are fluctuations in the number rather than the energy density so that
$\delta (n/s) \ne 0$. Examples of particle physics models giving
rise to isocurvature fluctuations include
axions \cite{axion,burns}, inflation with more than one inflaton field
\cite{infla}, and supersymmetric theories with flat directions
in the potential \cite{johncmb}.

Isocurvature perturbations\cite{EB86,general} in a given particle species 
have $\delta\rho=0$ with the overdensities
balanced by perturbations in other particle species, such as radiation.
At the last scattering surface (LSS) the compensation for the 
isocurvature perturbations can be maintained only for scales larger
than the horizon, effectively generating extra power to photon
perturbations at small angular multipoles $l$. As a consequence,
the spectrum of isocurvature perturbations differs a great deal
from adiabatic perturbations, and a purely isocurvature cold
dark matter (CDM) perturbation
spectrum is in fact already ruled out \cite{ruleout}
on the basis of COBE normalization
and $\sigma_8$, the 
amplitude of the rms mass fluctuations in an $8h^{-1}$ Mpc$^{-1}$
sphere. (There are suggestions that a decaying CDM model could 
sustain purely isocurvature perturbations \cite{hu}; however, such
models are not motivated from particle physics point of view). 
A small isocurvature contribution might,  however, be beneficial
 in the sense that it could help to improve the
fit to the power spectrum in $\Omega_{0} = 1$ 
CDM models with a cosmological constant \cite{SBG96,axion}.

The simplest possibility is that there is a single CDM which has both
adiabatic and isocurvature perturbations. An example of such a situation
is obtained in supersymmetric models with the so-called Affleck-Dine (AD) 
fields \cite{ad}, which also provide a basis
for baryogenesis. The AD fields are superpositions of squark and slepton
fields corresponding to the flat directions. During inflation they
will fluctuate along the flat directions  forming a condensate.
They are complex fields and, in the currently favoured D-term inflation
models \cite{dti},
is effectively massless during inflation.
The condensate does not correspond to the state of lowest energy
but fragments into non-topological solitons \cite{ks2}, 
which carry baryonic (and sometimes leptonic) number  \cite{cole2} and are
called B-balls (L-balls). The fluctuations in the AD field are both
adiabatic and isocurvature \cite{johncmb}, and will be inherited by
the B-balls, which are either stable or decay later into lightest 
supersymmetric particles,
which form the CDM. In the latter case there will be a lower bound on
the amplitude of the isocurvature perturbation.

Adiabatic perturbations\cite{KolbTurner,MFB92,LL93}
are characterized by the gauge-invariant
quantity $\zeta$, which equals $\delta\rho/(\rho + p)$ when
the perturbation exits or enters the
horizon.  Outside the horizon, $\zeta$ does not change with time.
Isocurvature perturbations\cite{EB86,MFB92,LL93} are characterized
by the gauge-invariant
entropy perturbation $ S \equiv \delta(n_c/s_\gamma) = \delta_c -
(3/4)\delta_\gamma $, where $n_c$ is the number density of CDM,
$s_\gamma$ is the entropy density associated with photons, and
$\delta_c$ and $\delta_\gamma$ are the relative overdensities in the CDM
and photon energy densities.  During radiation domination
$S \sim \delta_c \equiv \delta n_c/n_c$.
Outside the horizon, $S$ does not change with
time.  ``Initial'' scale-free ($n = 1$) adiabatic and isocurvature
perturbation spectra are given by
\bea
   P_\zeta(k) \equiv \langle|\zeta_k|^2\rangle = Ak^{n-4} = Ak^{-3},
   \nonumber \\
   P_S(k) \equiv \langle|S_k|^2\rangle = Bk^{n-4} = Bk^{-3}.
\eea

Following these remarks,  we assume in this paper
that there is just a single species of CDM, and
that it has both adiabatic and isocurvature fluctuations. As is natural
when both fluctuations have their origin in inflation, we also assume
that their power spectra are similar. The goal of the paper is then
to study how well particle physics
models with isocurvature fluctuations can be tested by the future
CMB satellite experiments, and at what level isocurvature perturbation
can be detected. In Sec.~II we define the 
Fisher information matrix needed for the cosmological parameter error 
estimates, and describe
the expected performance of MAP and Planck. In Sec.~III we discuss 
parametrization of isocurvature fluctuation and possible cosmological
models. In Sec.~IV we present the results of a numerical study for
MAP and Planck, while Sec.~V contains a discussion of the results.

\section{Accuracy of Parameter Determination}

   For a CMB temperature map with Gaussian fluctuations, all statistical
information is contained in the angular power spectrum
\beq
  C_{Tl} \equiv \langle|a^T_{lm}|^2\rangle.
\eeq
This spectrum contains a wealth of cosmological information.  The
major cosmological parameters can be determined from a high-quality
measurement of the spectrum, with an unprecedented accuracy.  This is
because the CMB anisotropy reflects conditions at the early universe, as
well as the geometry out to the present horizon, the microwave
photons having travelled through essentially the whole observable
universe as well as most of its history\cite{HuSu}.  There are however,
degeneracies with respect to some cosmological parameters; different
effects causing similar features in the power spectrum.  Also the
information from the large angular scales is severely limited by cosmic
variance.

Important additional information can be obtained by the measurement of
the polarization of the CMB\cite{ZS97,KKS97,HW97,HuWhite}.
Since polarization of the
CMB is caused by scattering only, polarization carries information about
the LSS only (including late scattering due to reionization),
eliminating confusion with line-of-sight effects.
Polarization is especially useful in separating out the effects of
tensor fluctuations, isocurvature fluctuations, and reionization, all of
which produce excess power at large scales. 

Unfortunately the polarization is expected to be of rather small
amplitude, at least an order of magnitude below the anisotropy signal,
making it difficult to detect\cite{KKS97}.
The contamination with the foreground
may also be a worse problem than for the temperature anisotropy, since
there are polarized foreground sources, e.g., galactic synchrotron
emission.  The MAP satellite is expected to provide a definite detection
of CMB polarization.  The higher sensitivity of the Planck Surveyor will
then help in getting a reasonable power spectrum for the polarization.

\begin{figure}
\vspace*{-2.5cm}
\epsfysize=13.0cm
\epsffile{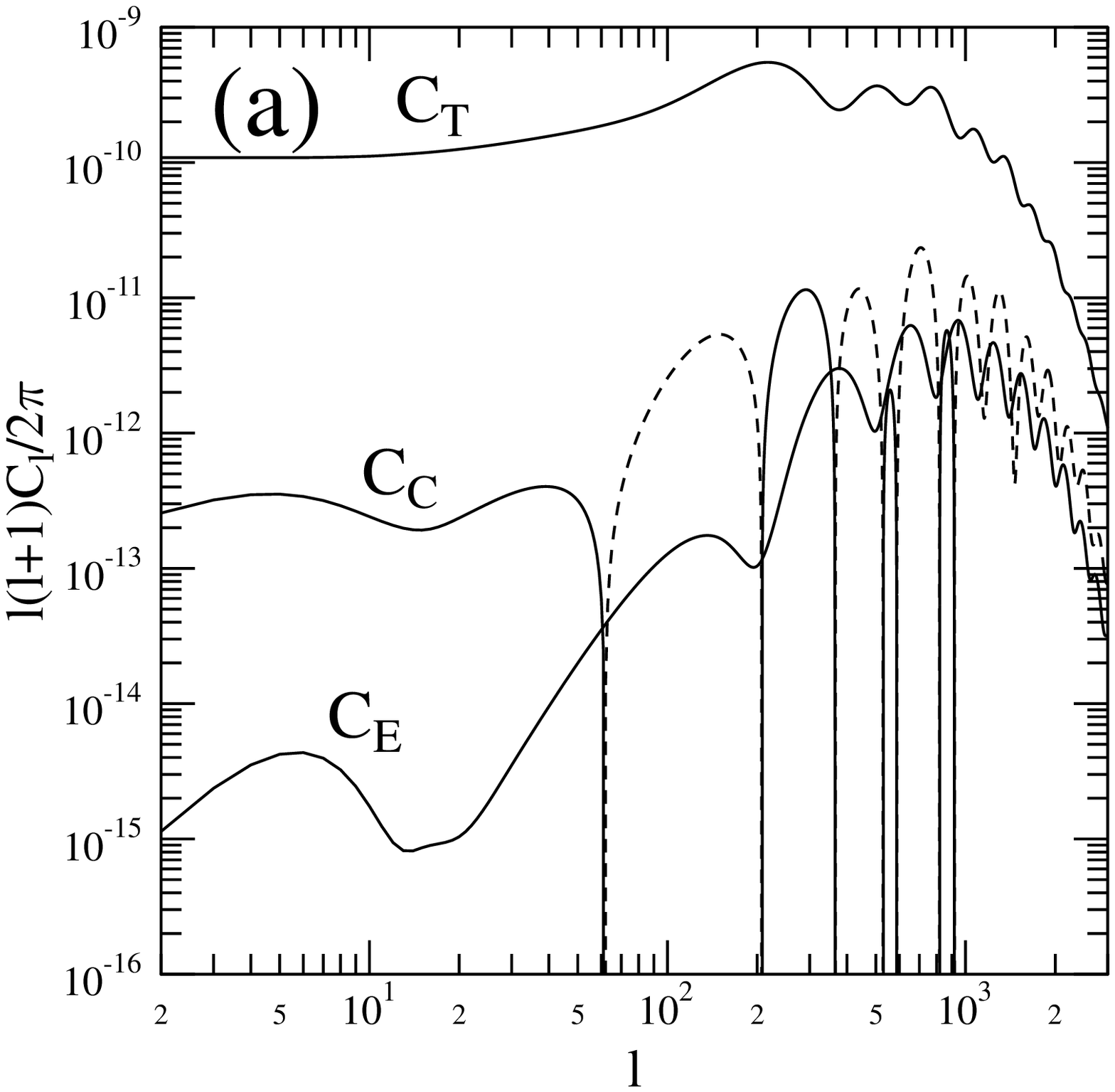}
\vspace*{-5.0cm}
\epsfysize=13.0cm
\epsffile{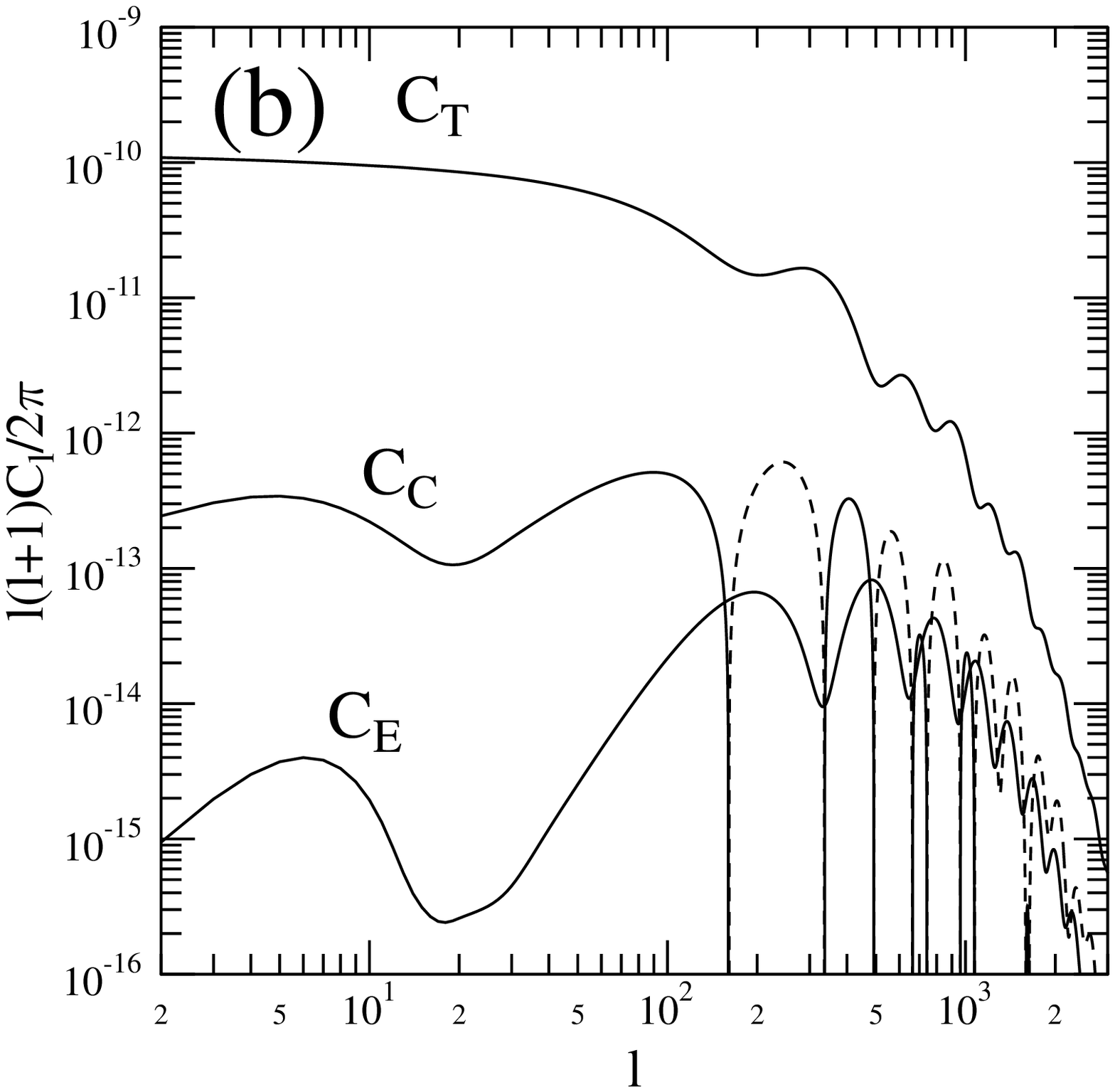}
\vspace*{-2.5cm}
\caption[a]{\protect
The angular power spectra for SCDM (see Table II) adiabatic (a) and
isocurvature (b) perturbations.  These scalar perturbations produce E-mode
polarization only, thus no B-mode spectra are shown.  The dashed lines
indicate negative values (for $C_{Cl}$).

}
\label{spectra}
\end{figure}

The polarization pattern on the sky can be separated into two
components, the E-mode and the B-mode\cite{ZS97}.
Thus an experiment which also
measures the polarization, will produce 
three maps: temperature,
E-mode polarization, and B-mode polarization.  The B-mode polarization
of the CMB is
uncorrelated with the temperature or the E-mode polarization.  Thus the
statistical information is carried by three angular power spectra
\bea
  C_{Tl} & \equiv & \langle|a^T_{lm}|^2\rangle, \nonumber \\
  C_{El} & \equiv & \langle|a^E_{lm}|^2\rangle, \nonumber \\
  C_{Bl} & \equiv & \langle|a^B_{lm}|^2\rangle,
\eea
and one correlation function
\beq
  C_{Cl} \equiv \langle a^{T*}_{lm}a^E_{lm}\rangle,
\eeq
giving the correlation between the temperature and the E-mode
polarization\cite{ZS97,KKS97,HW97,HuWhite,Kinney}.
The different spectra are demonstrated in Fig.~\ref{spectra}.

These four spectra can be calculated for different values of the
cosmological parameters ${s_i}$ and compared to data
from CMB experiments.  This way the true values of the cosmological
parameters can be determined.

Assuming the actual values of the parameters are ${\bar{s}_i}$, the
information obtained on these parameters by observation is described by
the Fisher information 
matrix\cite{Jungmanetal,TTH97,ZS97,Bondetal,ZSS97,Kinney}
\beq
  F_{ij} = \sum_l\sum_{X,Y}
  \frac{\partial C_{Xl}}{\partial s_i}|_{\bar{s}_i}
  \hbox{Cov}^{-1}(C_{Xl},C_{Yl})
  \frac{\partial C_{Yl}}{\partial s_j}|_{\bar{s}_j},
\eeq
where $C_{Xl} = C_{Tl}$, $C_{El}$, $C_{Bl}$, or $C_{Cl}$.

The covariance matrices, whose nonzero components are
\bea
  \hbox{Cov}(C_{Tl}C_{Tl}) & = & N_l\Biggl(C_{Tl}
  + \biggl[\sum_c w_T^cB_{l,c}^2\biggr]^{-1}\Biggr)^2, \nonumber\\
  \hbox{Cov}(C_{El}C_{El}) & = & N_l\Biggl(C_{El}
  + \biggl[\sum_c w_P^cB_{l,c}^2\biggr]^{-1}\Biggr)^2, \nonumber\\
  \hbox{Cov}(C_{Bl}C_{Bl}) & = & N_l\Biggl(C_{Bl}
  + \biggl[\sum_c w_P^cB_{l,c}^2\biggr]^{-1}\Biggr)^2, \nonumber\\
  \hbox{Cov}(C_{Cl}C_{Cl}) & = & N_l\Biggl[C_{Cl}^2
  +\Biggl(C_{Tl}+\biggl[\sum_c w_T^cB_{l,c}^2\biggr]^{-1}\Biggr) \times
  \nonumber\\
  & \times & \Biggl(C_{El}+\biggl[\sum_c w_P^cB_{l,c}^2\biggr]^{-1}\Biggr)
  \Biggr],  \nonumber\\
  \hbox{Cov}(C_{Tl}C_{El}) & = & N_lC_{Cl}^2, \nonumber\\
  \hbox{Cov}(C_{Tl}C_{Cl}) & = & N_lC_{Cl}\Biggl(C_{Tl}
  + \biggl[\sum_c w_T^cB_{l,c}^2\biggr]^{-1}\Biggr), \nonumber\\
  \hbox{Cov}(C_{El}C_{El}) & = & N_lC_{Cl}\Biggl(C_{El}
  + \biggl[\sum_c w_P^cB_{l,c}^2\biggr]^{-1}\Biggr),
\eea
depend on the sensitivity and angular resolution of the instrument
with 
\bea
  w_T^c = \frac{1}{\sigma_{T,c}^2\fwhmc^2}, \nonumber\\
  w_P^c = \frac{1}{\sigma_{P,c}^2\fwhmc^2},
\eea
and
\beq
  B_{l,c}^2 = e^{-l(l+1)\fwhmc^2/8\ln2}
\eeq
where $\sigma_c$ is the pixel noise (sensitivity) and $\fwhmc$ is
the beamwidth
(full width at half maximum) of instrument channel $c$.

The prefactor is given by
\beq
N_l=\frac{2}{(2l+1)\fsky},
\eeq
where $\fsky$ takes into account the fact that the whole sky cannot
be mapped because of 
foreground contamination. We adopt $\fsky = 0.65$, i.e., assume that
foreground can be completely removed from 65\% of the sky and the rest
of the sky is unusable.

The Planck Surveyor carries two instruments, HFI and LFI.
For MAP and LFI we use the three highest frequency channels, for HFI the
three lowest frequency channels for our estimates (see Table I).

\begin{table}
\begin{tabular}{lcccc}
instrument & frequency & $\fwhm$ & $\sigma_T$ & $\sigma_P$ \\ \hline
  MAP & 40 GHz & $0.47^\circ$ & 35 $\mu$K & $\sqrt{2}\times35\mu$K \\
      & 60 GHz & $0.35^\circ$ & 35 $\mu$K & $\sqrt{2}\times35\mu$K \\
      & 90 GHz & $0.21^\circ$ & 35 $\mu$K & $\sqrt{2}\times35\mu$K \\
  LFI & 44 GHz & 23' & $2.4\times10^{-6}$ & $\sqrt{2}\times2.4\times10^{-6}$ \\
      & 70 GHz & 14' & $3.6\times10^{-6}$ & $\sqrt{2}\times3.6\times10^{-6}$ \\
      &100 GHz & 10' & $4.3\times10^{-6}$ & $\sqrt{2}\times4.3\times10^{-6}$ \\
  HFI &100 GHz & 10.7' & $1.7\times10^{-6}$ & --- \\
      &143 GHz &  8.0' & $2.0\times10^{-6}$ & $3.7\times10^{-6}$ \\
      &217 GHz &  5.5' & $4.3\times10^{-6}$ & $8.9\times10^{-6}$ \\
\end{tabular}
\caption[a]{\protect 
The beamwidths and sensitivities of the satellite
experiments\cite{MAP,Planck}.}
\end{table}

The 1-$\sigma$ error on parameter $s_i$ is
\beq
  \Delta s_i = \sqrt{\bigl(F^{-1}\bigr)_{ii}},
\eeq
when all parameters are estimated from the same data.  (If all the other
parameters were somehow known a priori, the error would be $\Delta s_i =
(F_{ii})^{-\half}$).  This is a lower limit for the error; but when the
error is small it is usually a good approximation for the
error\cite{TTH97,ZSS97}.

\section{The Parameters and Models}

We consider a simultaneous fit of 10 parameters, $C_2$, $\Omega_0 \equiv
\Omega_\Lambda + \Omega_m$, $\Omega_\Lambda - \Omega_m$, $\Omega_b h^2$,
$\Omega_{\rm HDM}$, $H_0$, $\tau$, $n_S$, $r$, and $\alpha$.
Here $\tau$ is the optical depth to LSS and $r \equiv C_2^T/C_2^S$ is
the tensor/scalar ratio.  We assume three matter components,
$\Omega_m = \Omega_{\rm CDM} + \Omega_{\rm HDM} + \Omega_b$.  Note that
we have defined $\Omega_0$ to include the vacuum energy; it is the
parameter which determines the curvature of space.  
We assume $T_0 = 2.728$ K, $Y_{\rm
He} = 0.24$, and $N_\nu = 3$.  For hot dark matter we assume all 3
neutrinos have equal masses.

The isocurvature contribution to anisotropy is given by the parameter
$\alpha$, which gives the fraction of $C_2^S$ due to isocurvature
perturbations.
Thus
\beq
  \alpha \equiv C_2^{\rm iso}/(C_2^{\rm iso}+C_2^{\rm ad}).
\eeq
Our $\alpha$ is the same as $1-\alpha$ of Stompor et al\cite{SBG96}. 
The isocurvature fluctuations are assumed
to be primordial density fluctuations in CDM, initially balanced by
radiation+baryons+neutrinos to make the total energy density
homogeneous.  The isocurvature spectral index is assumed to be the same
as for adiabatic fluctuations; in principle, they could be different,
but if both fluctuations originate from inflation, and in particular
from massless field fluctuations during inflation, then their power spectra
are guaranteed to be equal.

The large-scale anisotropy due to isocurvature perurbations 
is\cite{LL93}
\beq
   \frac{\delta T}{T} = -\frac{1}{3} S + \frac{1}{3} \Phi,
\eeq
where 
\beq
   \Phi = -\frac{1}{5} S
\eeq
is the Newtonian gravitational
potential\cite{Bardeen,MFB92,HuSu},
whereas for adiabatic perturbations we have
just
\begin{table}
\begin{tabular}{llllllllll}
 & $\Omega_0$ & $\Omega_\Lambda$ & $\Omega_b$ & $\Omega_{\rm HDM}$ 
 & $H_0$ & $\tau$ & $n_S$ & $r$ & $\alpha$ \\ \hline
        SCDM & 1.0 & 0   & 0.05 & 0 & 50 & 0.05 & 1 & 0 & 0 \\
$\Lambda$CDM & 1.0 & 0.7 & 0.06 & 0 & 65 & 0.1  & 1 & 0 & 0 \\
        OCDM & 0.4 & 0   & 0.06 & 0 & 65 & 0.05 & 1 & 0 & 0 \\
\end{tabular}
\caption[a]{\protect The cosmological models considered.}
\end{table}
\beq
   \frac{\delta T}{T} = \frac{1}{3} \Phi,
\eeq
where the potential is related to the gauge-invariant $\zeta$ by
\beq
   \Phi = -\frac{3}{2}\zeta
\eeq

Thus the ``observational'' isocurvature parameter $\alpha$ is related 
to the amplitudes of the initial isocurvature and adiabatic spectra by
\beq
   \frac{\alpha}{1-\alpha} = \frac{(-\delta T/T)_{\bf iso}}
                                  {(-\delta T/T)_{\bf ad}}
   = \biggl(\frac{4S}{5\zeta}\biggr)^2 =
\frac{16}{25}\biggl(\frac{B}{A}\biggr).
\eeq

In a given particle physics model $A$ and $B$ are known in terms
of the model parameters, and Eq. (17) tells how these parameters
are related to the observed isocurvature parameter $\alpha$.
                        
We have chosen to consider three different reference
cosmological models (see Table II), which
are the same as discussed by  Wang et al.\cite{WSS99}: the standard 
CDM model with $\Omega_0=1$,
a CDM model with a cosmological constant $\Omega_\Lambda=0.7$ and
$\Omega_0=1$, and an open CDM model with $\Omega_0=0.4$.

To find the derivatives $\partial C_{Xl}/\partial s_i$ we used
CMBFAST\cite{CMBFAST} to run
variations around these three reference models, with steps $\Delta s_i$
or $\Delta s_i/s_i$ between 0.01 and 0.05.

We COBE-normalized\cite{BunnWhite} $C_2$ for the
three reference models.
Note that as $C_2$ is one of the parameters to be determined from
the data, we
do not COBE-renormalize it for the variations around these reference
models.

Wang et al.\cite{WSS99} considered 7-parameter fits.  Our
additional parameters are $\Omega_{\rm HDM}$, $r$, and $\alpha$.  For
OCDM we considered 8-parameter fits only,
keeping $r \equiv 0$ and $\Omega_{\rm
HDM} \equiv 0$, since CMBFAST does not allow nonzero values for these
parameters in open models.

\section{Results}

We are interested in how accurately the isocurvature contribution can be
determined and what is the smallest contribution that can still be
detected.
In the following tables (III, IV, and V)
we give the 1-$\sigma$ error on the isocurvature
parameter $\alpha$ for the three experiments, MAP, Planck LFI, and Planck
HFI, and an ideal experiment with zero noise and infinitely fine
resolution.  Full sky coverage would reduce these errors by a factor of
$\sqrt{0.65}$.

We see that the addition of the two parameters, $\Omega_{\rm HDM}$ and $r$,
increased the error in $\alpha$ determination significantly for the case
without polarization data, but had only a small effect for the case with
polarization.  This is because the temperature spectra from tensor and
isocurvature fluctuations resemble each other, resulting in a
degeneracy
between the parameters $r$ and $\alpha$.  (Adding $\Omega_{\rm HDM}$ as a
free parameter changes $\Delta\alpha$ at most by 10\%.)  Because only
tensor fluctuations have B-mode polarization, polarization measurements
break this degeneracy.

\begin{figure}
\vspace*{-2.7cm}
\epsfysize=13.0cm
\epsffile{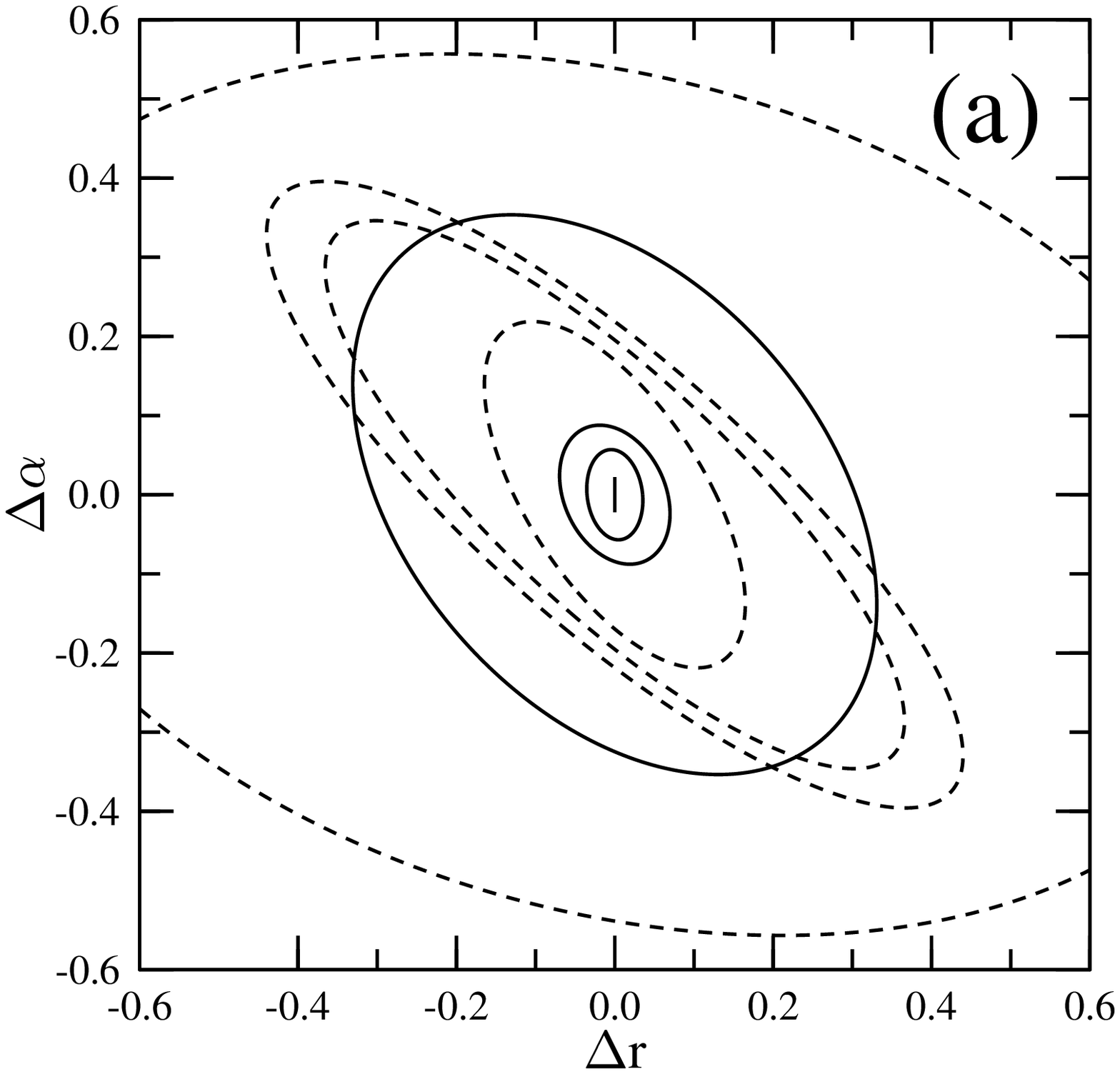}
\vspace*{-5.2cm}
\epsfysize=13.0cm
\epsffile{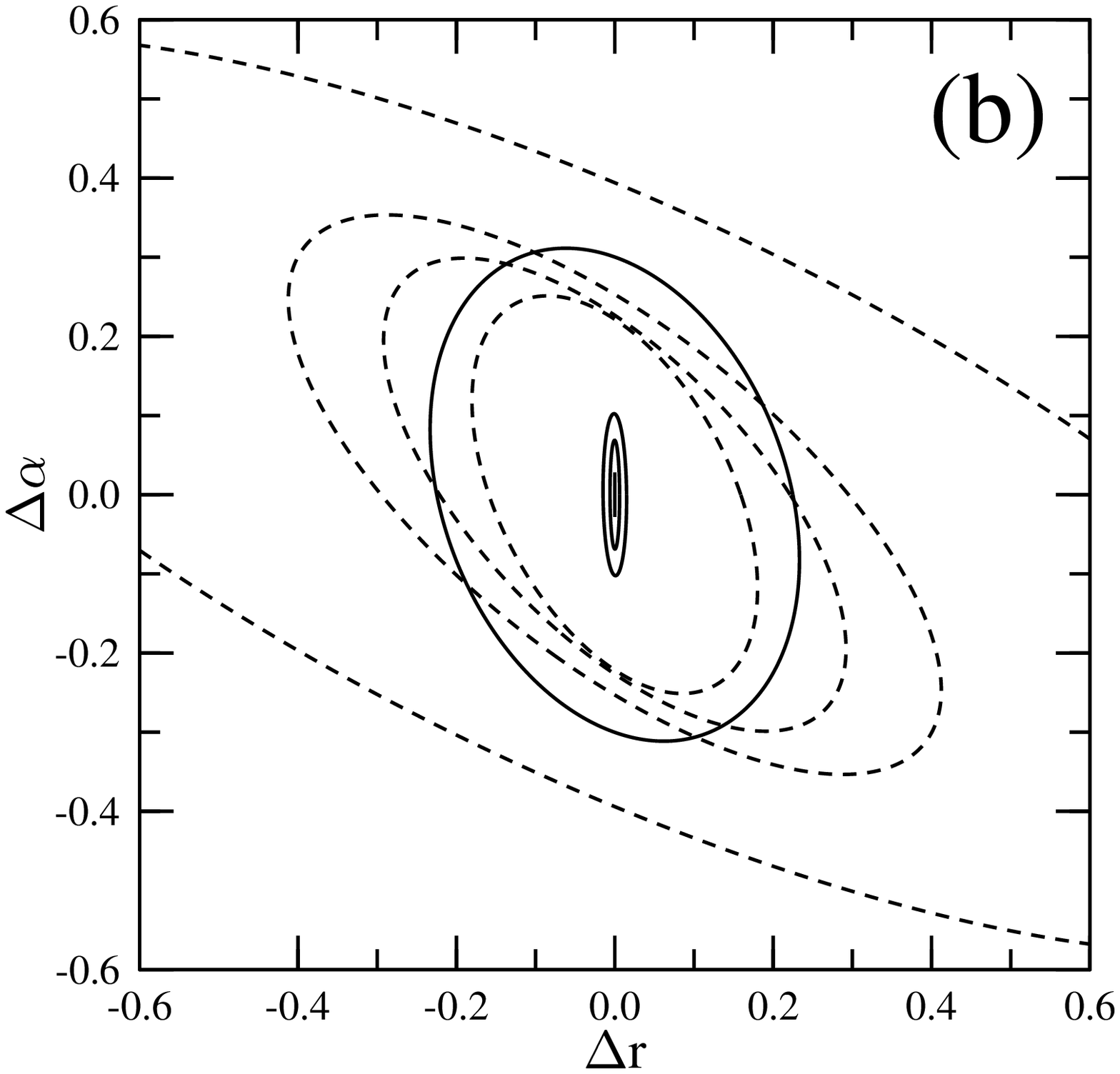}
\vspace*{-2.7cm}
\caption[a]{\protect
The error ellipses in the ($r$,$\alpha$)-plane for (a) SCDM and (b)
$\Lambda$CDM. The dashed lines are
for the case without polarization measurement, the solid lines with
polarization.  The four ellipses from the outside inward are for MAP,
Planck LFI, Planck HFI, and the ideal experiment.  
}
\label{ellipses}
\end{figure}

\begin{table}
\begin{tabular}{lllll}
 & MAP & LFI & HFI & ideal \\ \hline
\multicolumn{5}{c}{Without polarization} \\ \hline
SCDM         & 0.37   & 0.26   & 0.23   & 0.14   \\
$\Lambda$CDM & 0.38   & 0.23   & 0.20   & 0.17   \\ \hline
\multicolumn{5}{c}{With polarization} \\ \hline
SCDM         & 0.23   & 0.058  & 0.038  & 0.015  \\
$\Lambda$CDM & 0.21   & 0.067  & 0.045  & 0.019  \\
\end{tabular}
\caption[a]{\protect $\Delta\alpha$ from 10-parameter fits.}
\end{table}

Note that the tensor and isocurvature parameters, $r$ and $\alpha$ are
non-negative by definition.
It is quite likely that their true values
are smaller than the errors in their determination, so that
we will only obtain an upper limit.
For the case without polarization information,
there will be a degeneracy between these two parameters, such that their
sum can be obtained with a higher accuracy than the parameters
themselves (see Fig.~\ref{ellipses}).
Since the upper limit to the sum is also an upper
limit to the individual parameters, we can then obtain a tighter upper limit
to $\alpha$ than the $\Delta\alpha$ given in Table III.  This upper limit
is essentially the $\Delta\alpha$ given in Table IV, where we have made
the restriction $r\equiv0$.

For MAP, there are also other significant degeneracies among the
10 parameters considered here.
For Planck without polarization, the only other important 
degeneracy is with the optical depth $\tau$ due to reionization.
Polarization breaks this degeneracy, since early reionization produces
a polarization signal at large angular scales which cannot be mimicked
by other parameter combinations\cite{Zaldarriaga97}.
For Planck with polarization there are no
significant degeneracies
between $\alpha$ and the other parameters (except, of course,
a degeneracy between $C_2$ and $\alpha$),
and so the errors would not
become much smaller even if all the other parameters were assumed known
(see Table V).

\begin{table}
\begin{tabular}{lllll}
 & MAP & LFI & HFI & ideal \\ \hline
\multicolumn{5}{c}{Without polarization} \\ \hline
SCDM         & 0.35   & 0.13   & 0.12   & 0.10   \\
$\Lambda$CDM & 0.22   & 0.16   & 0.15   & 0.15   \\
OCDM         & 0.27   & 0.13   & 0.064  & 0.053  \\ \hline
\multicolumn{5}{c}{With polarization} \\ \hline
SCDM         & 0.21   & 0.056  & 0.037  & 0.015  \\
$\Lambda$CDM & 0.16   & 0.062  & 0.044  & 0.018  \\
OCDM         & 0.14   & 0.046  & 0.033  & 0.012  \\
\end{tabular}
\caption[a]{\protect $\Delta\alpha$ from 8-parameter fits ($\Omega_{\rm
HDM} \equiv 0$, $r \equiv 0$).}
\end{table}

\begin{table}
\begin{tabular}{lllll}
 & MAP & LFI & HFI & ideal \\ \hline
\multicolumn{5}{c}{Without polarization} \\ \hline
SCDM         & 0.053  & 0.049  & 0.047  & 0.046  \\
$\Lambda$CDM & 0.055  & 0.052  & 0.051  & 0.050  \\
OCDM         & 0.041  & 0.038  & 0.036  & 0.035  \\ \hline
\multicolumn{5}{c}{With polarization} \\ \hline
SCDM         & 0.053  & 0.039  & 0.030  & 0.013  \\
$\Lambda$CDM & 0.054  & 0.044  & 0.036  & 0.016  \\
OCDM         & 0.041  & 0.030  & 0.026  & 0.010  \\
\end{tabular}
\caption[a]{\protect $\Delta\alpha$ from 2-parameter fits ($C_2$ and
$\alpha$).  This table is shown to illustrate, by comparison to Table
IV, the effect of degeneracies with other parameters.  Because the best
determinations of many of the other parameters have to come from the
same CMB data, $\alpha$ cannot in reality be determined with this
accuracy.}
\end{table}

\section{Discussion}

{}From Tables III and IV we find that the (1-$\sigma$) detection limit
of isocurvature fluctuations
is $\alpha \sim 0.04$
for the Planck Surveyor.  This corresponds to a ratio
$B/A \sim 0.07$ between the amplitudes of the initial isocurvature
and adiabatic perturbation spectra. This is at the level
predicted by certain B-ball models \cite{johncmb},
which thus could be tested by Planck. It is likely that other particle 
physics models can also be constrained in a significant way on the
basis (or the absence) of isocurvature fluctuation in CMB. However, 
to reach this sensitivity polarization measurements are crucial,
as is evident from Tables III and IV. 

Without polarization information, the contribution from isocurvature
fluctuations to temperature anisotropies could be confused with tensor
perturbations or early reionization effects. 
Polarization measurements break this degeneracy, since these other two
effects both have a unique polarization signal.

Whether such sensitivity can be realized is still unclear; the foreground
contamination for the polarization spectra
may turn out to be a more severe problem
than in the case of the temperature angular power spectrum. 
However, optimizing the polarization information should have a 
high priority, in particular for Planck; because of its higher
sensitivity, it is in this respect that Planck is truly a superior
instrument compared with MAP. CMB polarization measures both
tensor and isocurvature contribution, which are of paramount
interest to inflation model builders.

Finally, the
cosmic variance sets an absolute lower theoretical limit, 
at about $\alpha \sim 0.01$, to
how small isocurvature fluctuations can be detected from the CMB
anisotropy by any experiment.

\section*{Acknowledgements}

This work has been supported by the
 Academy of Finland  under the contract 101-35224.
We thank
the Center for Scientific Computing (Finland) for computational resources.
We acknowledge the use of the CMBFAST Boltzmann code developed by
Uro\v{s} Seljak and Matias Zaldarriaga.

\end{document}